\documentclass{article}
\usepackage{microtype}
\usepackage[dvipsnames]{xcolor}
\usepackage{graphicx}
\usepackage{booktabs} 
\usepackage{multirow}
\usepackage{vcell}
\DeclareTextSymbolDefault{\dh}{T1}
\usepackage{hyperref}

\usepackage[accepted]{icml2021}

\usepackage{multirow}
\usepackage{vcell}
\usepackage{subcaption}
\usepackage{caption}
\usepackage{subcaption}
\usepackage{verbatim}
\usepackage{subcaption}

\icmltitlerunning{SALT: \textbf{S}ea lice \textbf{A}daptive \textbf{L}attice \textbf{T}racking}

\begin{document}

\twocolumn[
    \icmltitle{SALT: \textbf{S}ea lice \textbf{A}daptive \textbf{L}attice \textbf{T}racking - An Unsupervised\\Approach to Generate an Improved Ocean Model}
    
    \icmlsetsymbol{equal}{*}
    
    \begin{icmlauthorlist}
        \icmlauthor{Ju An Park}{blue}
        \icmlauthor{Vikram Voleti}{blue,mila,mon}
        \icmlauthor{Kathryn E. Thomas}{blue,otu}
        \icmlauthor{Alexander Wong}{blue,wat}
        \icmlauthor{Jason L. Deglint}{blue,wat}
        \vspace{-0.25cm}
    \end{icmlauthorlist}
    
    \icmlaffiliation{blue}{Blue Lion Labs, Waterloo, Ontario, Canada}
    \icmlaffiliation{otu}{Ontario Tech University, Ontario, Canada}
    \icmlaffiliation{mila}{Mila, Montreal, Canada}
    \icmlaffiliation{mon}{Universit\'e de Montr\'eal, Montreal, Canada}
    \icmlaffiliation{wat}{University of Waterloo, Waterloo, Canada}
    
    \icmlcorrespondingauthor{Ju An Park}{juan@bluelionlabs.com}
    
    \icmlkeywords{dispersion modelling, CCAI, sea lice, particle tracking, adaptive mesh, stochastic modelling}
    
    \vskip 0.3in
]



\printAffiliationsAndNotice{}  

\begin{abstract}
Warming oceans due to climate change are leading to increased numbers of ectoparasitic copepods, also known as sea lice, which can cause significant ecological loss to wild salmon populations and major economic loss to aquaculture sites. The main transport mechanism driving the spread of sea lice populations are near-surface ocean currents. Present strategies to estimate the distribution of sea lice larvae are computationally complex and limit full-scale analysis. Motivated to address this challenge, we propose SALT: Sea lice Adaptive Lattice Tracking approach for efficient estimation of sea lice dispersion and distribution in space and time. Specifically, an adaptive spatial mesh is generated by merging nodes in the lattice graph of the Ocean Model based on local ocean properties, thus enabling highly efficient graph representation. SALT demonstrates improved efficiency while maintaining consistent results with the standard method, using near-surface current data for Hardangerfjord, Norway. The proposed SALT technique shows promise for enhancing proactive aquaculture management through predictive modelling of sea lice infestation pressure maps in a changing climate.
\vspace{-0.6cm}
\end{abstract}

\section{Introduction}

Sea lice (\textit{Lepeophtheirus salmonis and Caligus} sp.), are marine ectoparasitic copepods that often have outbreaks on salmon farms causing significant environmental and economical damage \cite{ford_global_2008, torrissen_salmon_2013, wild_salmon_damage}. Two contributing factors in the increased risk and impacts of sea lice outbreaks are (1) warming ocean temperatures and (2) the increasing scale of aquaculture. First, warming oceans induced by climate change \cite{abraham_review_2013} speed up the life cycle of sea lice \cite{pike_sealice_1999}, allowing them to persist in larger numbers for a longer duration \cite{costello_ecology_2006, wild_salmon_damage_due_to_control_failure, warming_decreases_salmon_survival}. Second, salmon in aquaculture farms are housed in high density, open-water cages, making them ideal breeding grounds for sea lice. As aquaculture has been providing more than double the wild-caught supply of salmon since 2012 \cite{mowi_salmon_2019}, and is only expected to increase with time, exacerbating the risk of sea lice infestations.

For effective mitigation and proactive management of sea lice, their populations need to be estimated across space and time. To accomplish this, infestation pressure (also known as spatial density) forecast models \cite{barrett_prevention_2020} are used. For example, Norway has implemented a sea lice traffic light system\footnote{https://www.hi.no/forskning/marine-data-forskningsdata/lakseluskart/html/lakseluskart.html} which uses an infestation pressure forecast model to inform management decisions, such as restricting farming in designated production zones to prevent sea lice infestations \cite{myksvoll_evaluation_2018}. Current state-of-the-art forecast models \cite{myksvoll_evaluation_2018} run on super computers to provide weekly sea lice pressure estimates for coastal regions. One limitation of these models is that they cannot estimate infestation pressures in local systems (e.g. small fjords), as a higher resolution Ocean Model is required. While finer resolution Ocean Models exist, they are not utilized due to bottlenecks in computation capacity. Thus, there is a direct benefit to improving the computational efficiency of current sea lice density estimation models to enable full-scale models capable of estimating pressures for both coast and local water systems.

Given the current limitations, we propose Sea lice Adaptive Lattice Tracking (SALT) to estimate sea lice particles with less storage requirements. 
This technique takes advantage of spatio-temporal patterns
that exist in standard Ocean Models to generate an adaptive Ocean Model, which is fed into a Dispersion Model to provide estimates of sea lice particles (see \autoref{fig:overview}). This adaptive Ocean Model reduces storage requirements while maintaining strong estimation accuracy compared to the standard density estimation models (see \autoref{table_ocean_model} and \autoref{fig_num_cluster}).
We hope that our improvements on these models will support sustainable fish farming in both coastal and local water systems in a changing climate.

\section{Standard Method (OSLM)}

\begin{figure}[!t]
    \centering
    \includegraphics[width=0.9\linewidth]{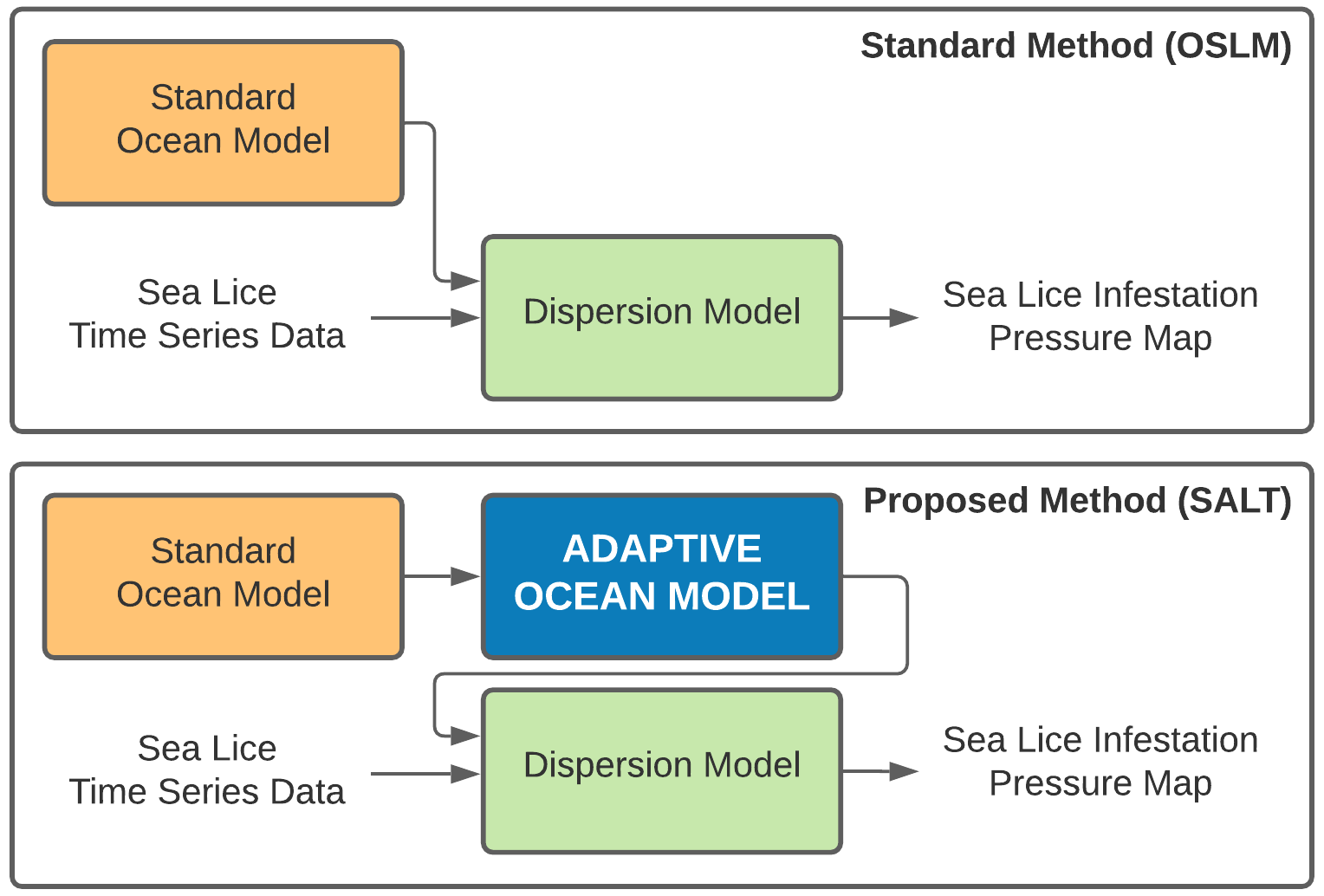}

    \caption{The standard method (top), Operational Salmon Lice Model (OSLM), consists of two inputs into the Dispersion Model: the standard Ocean Model, and the sea lice time series data. This Dispersion Model creates the sea lice infestation pressure map. Our proposed method (bottom) uses uses the standard Ocean Model to create an adaptive Ocean Model, which provides an efficient method to estimate sea lice particles with notably lower data storage requirements, while maintaining strong estimation accuracy compared to the standard approach.}
    \label{fig:overview}
\end{figure}

Here, we present the baseline particle tracking method used to forecast infestation pressures, which we will refer to as the Operational Salmon Lice Model (OSLM) method. This method is currently in use in Norway as part of the Traffic Light System \cite{myksvoll_evaluation_2018}. As seen in \autoref{fig:overview} (top), the main method for forecasting sea lice infestation pressures consists of three main components: (1) an Ocean Model, (2) sea lice time series data, and (3) Dispersion Model for particle tracking.

\textbf{Ocean Model:} The Ocean Model is a large spatio-temporal grid which describes ocean parameters including current speed, salinity, and temperature. It is represented by a lattice graph where each node in the graph represents the ocean parameters at a specific fixed square area. Storage complexity for these models are approximately $m \times x \times y \times z \times t$, where $m$ is the number of ocean parameters per node, $x$, $y$, and $z$ is the size of the grid along the horizontal and vertical spatial axis, and $t$ is the size of the grid along the temporal axis. Such a lattice graph representation can lead to significant computational complexities that make it very challenging for full-scale spatial estimation of sea lice dispersion and distribution. For reference, a single day's worth of data for the NorKyst 800m resolution grids uses 5.7 GB amount of storage \cite{asplin_hydrodynamic_2020}. Since storage requirements for these models scale quadratically with respect to resolution, it is important to improve their efficiency.

\textbf{Sea Lice Time Series Data} : This describes the average number of female sea lice per fish at aquaculture sites. This data is important as it allows the particle tracking models to generate realistic quantities of sea lice at each site during the simulation \cite{myksvoll_evaluation_2018}. 

\textbf{Dispersion Model:} The Dispersion Model takes the Ocean Model and sea lice time series data as input, simulates sea lice particles at different locations at the initial time step, and calculates the motion of each particle to create a \textbf{sea lice infestation pressure map}. These models have been used to understand sea lice trajectories in specific areas with a simulation period from a week to more than a month \cite{amundrud_modelling_2009}, and realistic sea lice pressures of entire coastal regions \cite{asplin_dispersion_2014,sandvik_toward_2016, myksvoll_evaluation_2018}. The motion of the particles is assumed to follow a lagrangian transport process solved using a Runge Kutta 4th order ordinary differential equation solver \cite{press_numerical_2007} with a 3 minutes time step. Furthermore, a random movement component is added to the particle motion to represent the uncertainty in physical processes finer than the Ocean Model resolution processes and sea lice movement behavior \cite{myksvoll_evaluation_2018}.

\textbf{Discussion:} This technique has been implemented in Norway to control sea lice levels in the form of a Traffic Light System, which has been proven to provide accurate estimations of sea lice concentrations for the coastal waters of Norway \cite{myksvoll_evaluation_2018}. However, the current implementation of the system is operational mostly in coastal waters, as the Ocean Model used is an 800 m grid, which is too coarse to model local systems. While higher resolution Ocean Models exist \cite{asplin_dispersion_2014}, the computational requirements yet remains too high for a full-scale implementation \cite{myksvoll_evaluation_2018}.

\section{Proposed Method (SALT)}

\begin{figure}[t]
     \centering
     \includegraphics[width=.8\linewidth]{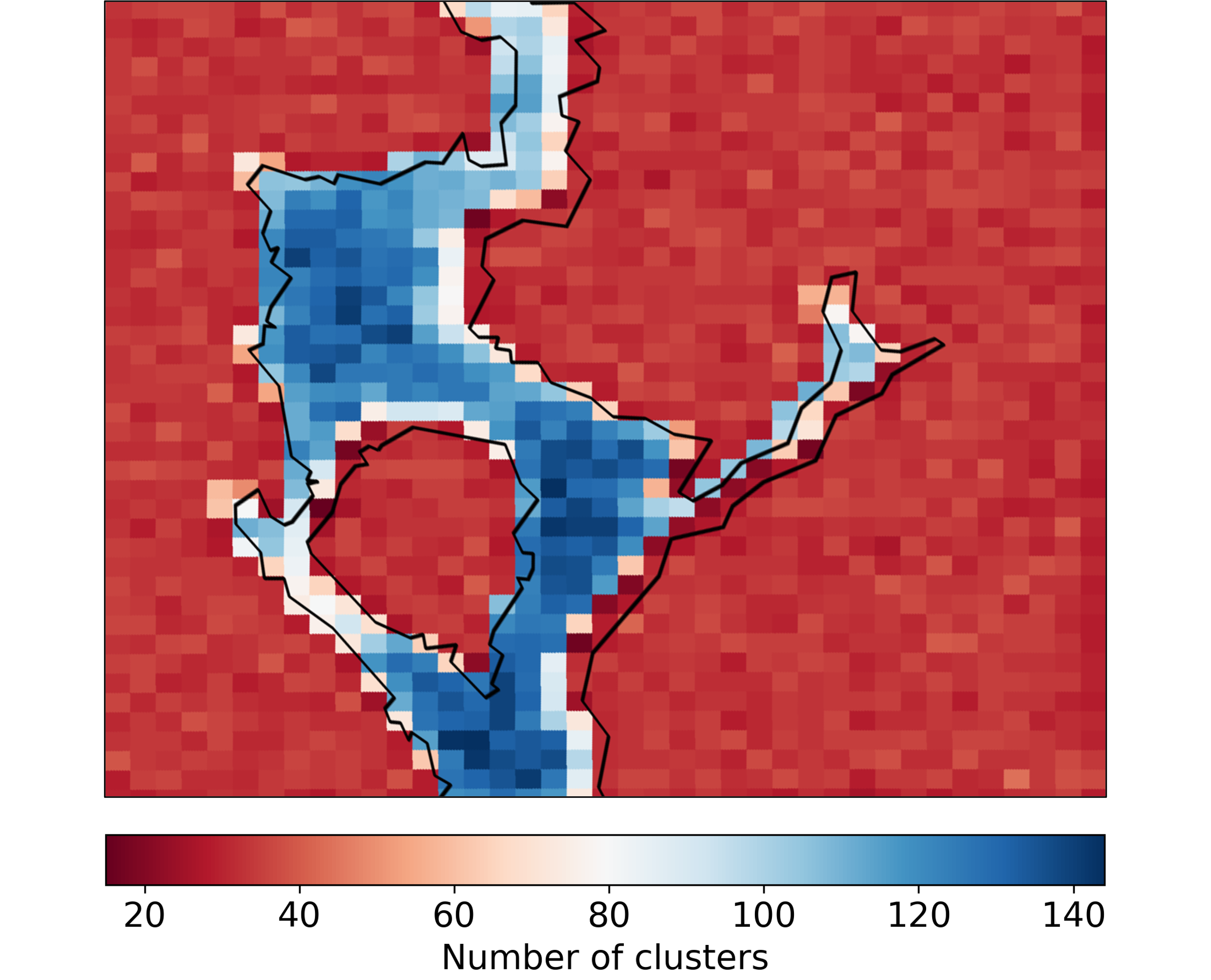}
    \caption{Map of the number of clusters associated with each  grid point over the course of the simulation. Low number of clusters in a grid point indicate high degree of temporal aggregation. Black lines are the approximate coastlines e.g. the boundary between land and water. It can be seen that areas on land which have less temporal change have significantly fewer clusters, while areas in water have more clusters.}
    \label{fig_num_cluster}
    \vspace{-1em}
\end{figure}

As seen in \autoref{fig:overview} (bottom), we introduce an adaptive spatio-temporal mesh grid to reduce the memory and storage requirements of the Ocean Model while maintaining performance. This adaptive Ocean Model grid is then input to the same Dispersion Model as the OSLM method.

\textbf{Adaptive Ocean Model:} Our adaptive Ocean Model is generated by merging nodes in the lattice graph of the standard Ocean Model through assignment of one or more nodes with similarities to a single super node (also known as clusters). These nodes are merged using k-means clustering by aggregating similar local parameters in the Ocean Model such as spatial and temporal coordinates, horizontal/vertical current, and a Boolean indicating whether the node is on land or water. The overall loss ($L$) for the k-means clustering is:

\begin{equation}
    L = \sum_{i=1}^p \sum_{j=1}^k ||x_i-c_k||
\end{equation}
\noindent where $p$, $k$, $x_i$, and $c_k$ respectively represent the total number of sea lice particles, the total number of clusters, the particle vector, and the cluster vector. The particle and cluster vector contain parameters relevant to clustering, such as spatial and temporal coordinates, and horizontal/vertical current speed.

Through this unsupervised clustering method, we are able to intelligently and autonomously compress the Ocean Model by aggregating areas of low spatio-temporal variability into super-nodes, while areas with high variability are left untouched. Note that the number of initialized super-nodes is equal to the number of nodes in the adaptive ocean model, meaning that the desirable number of nodes can be chosen.

\section{Experimental Setup}

To conduct a direct comparison between the OSLM method and our proposed SALT method, the same Ocean Model and sea lice generation scheme have been used. We use the hourly forecast NorKyst800m Ocean Model of the Hardangerfjord region from February 7, 2020 to February 14, 2020 \cite{asplin_hydrodynamic_2020}. We compare the final dataset size in megabytes and the number of nodes. Only the topmost surface ocean currents of NorKyst800m are used for sea lice dispersion modelling as sea lice behaviour studies have shown to date that sea lice are not likely to go below depths of 10m \cite{amundrud_modelling_2009}.

To evaluate the accuracy of SALT against OSLM, we chose to use the same sea lice generation scheme to ensure that all points in the grid were being well represented in the evaluation.
Specifically, 10 sea lice particles are randomly generated in each water node in 24 hour intervals beginning from the start of the Dispersion Model simulation.

We compare the Ocean Model in the standard method and our Adaptive Ocean Model in terms of number of total nodes, ocean model size, and add details of the additional time taken to perform k-means clustering for our method (see \autoref{table_ocean_model}). We then compare the root mean squared error (RMSE) of the infestation pressure maps forcasted out to seven days from both methods (\autoref{table_disp_model} and \autoref{fig_RMSE}). A workstation with an AMD Threadripper 3960X and 128 GB RAM was used for all computations.

\section{Results \& Discussion}

\begin{figure*}[!th]
     \centering
     \begin{subfigure}[b]{0.3\linewidth}
         \centering
         \includegraphics[width=\linewidth]{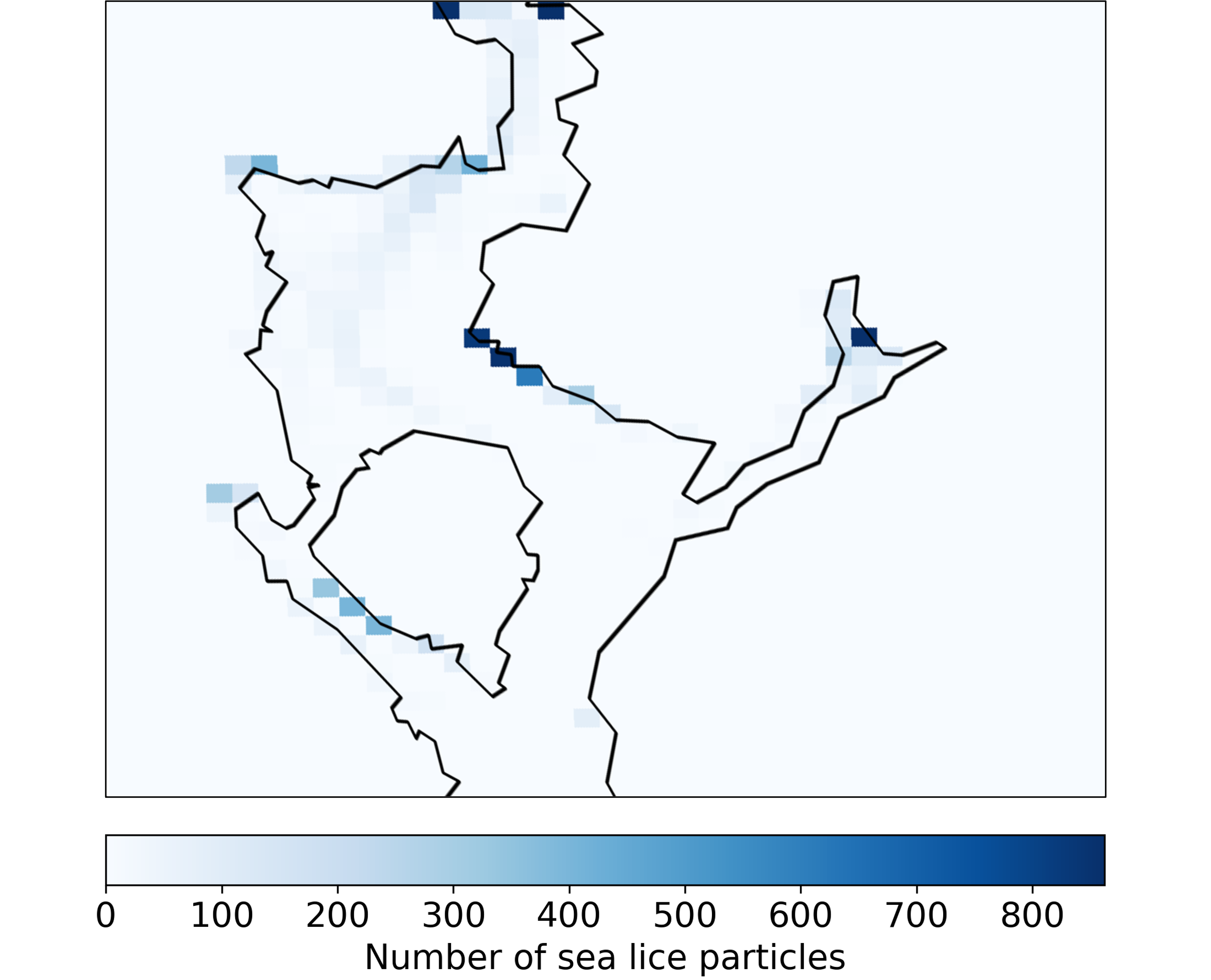}
         \caption{OSLM}
         \label{fig_OSLM}
     \end{subfigure}
     \hfill
     \begin{subfigure}[b]{0.3\linewidth}
         \centering
         \includegraphics[width=\linewidth]{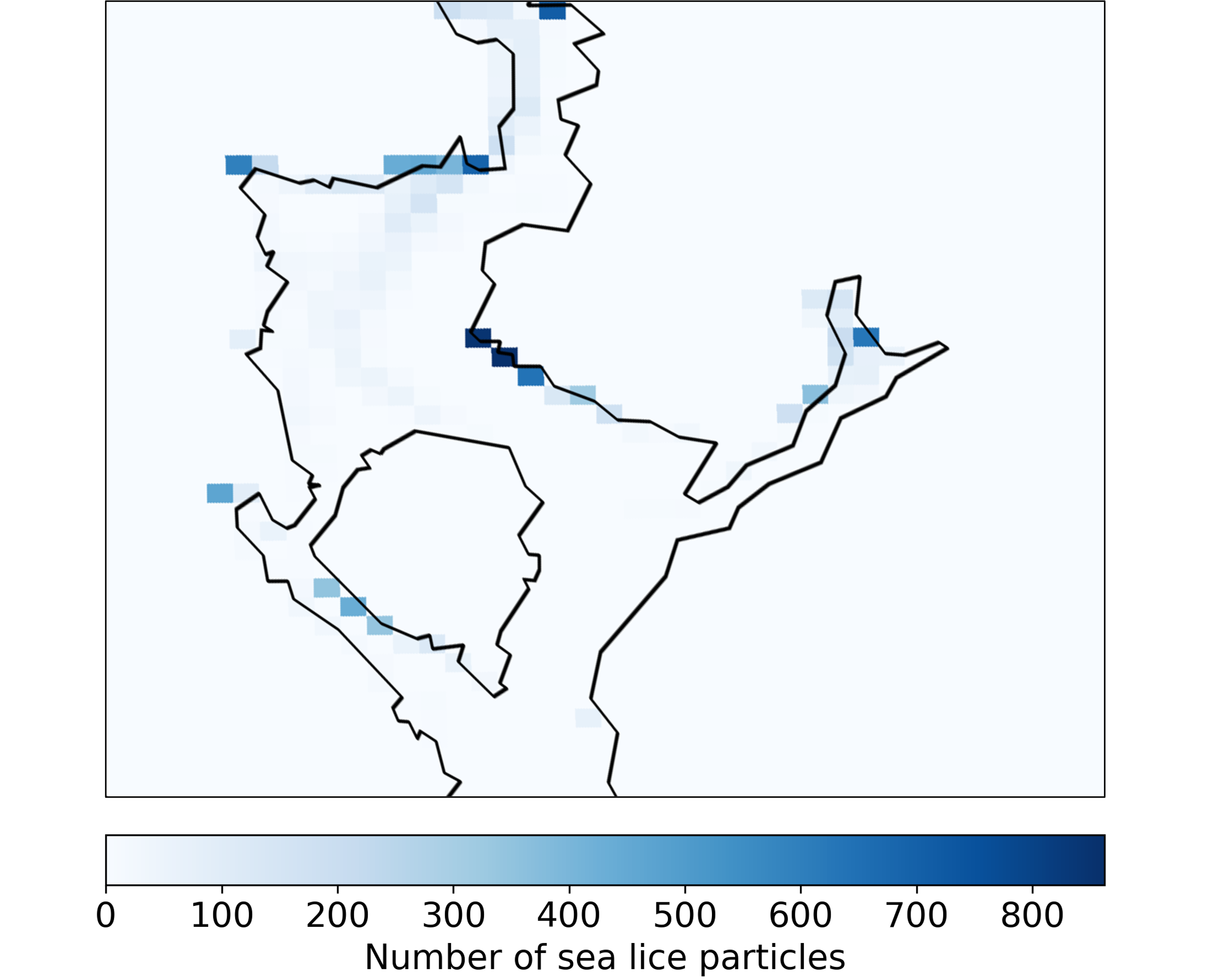}
         \caption{SALT (K=10\%)}
         \label{fig_SALT}
     \end{subfigure}
     \hfill
     \begin{subfigure}[b]{0.3\linewidth}
         \centering
         \includegraphics[width=\linewidth]{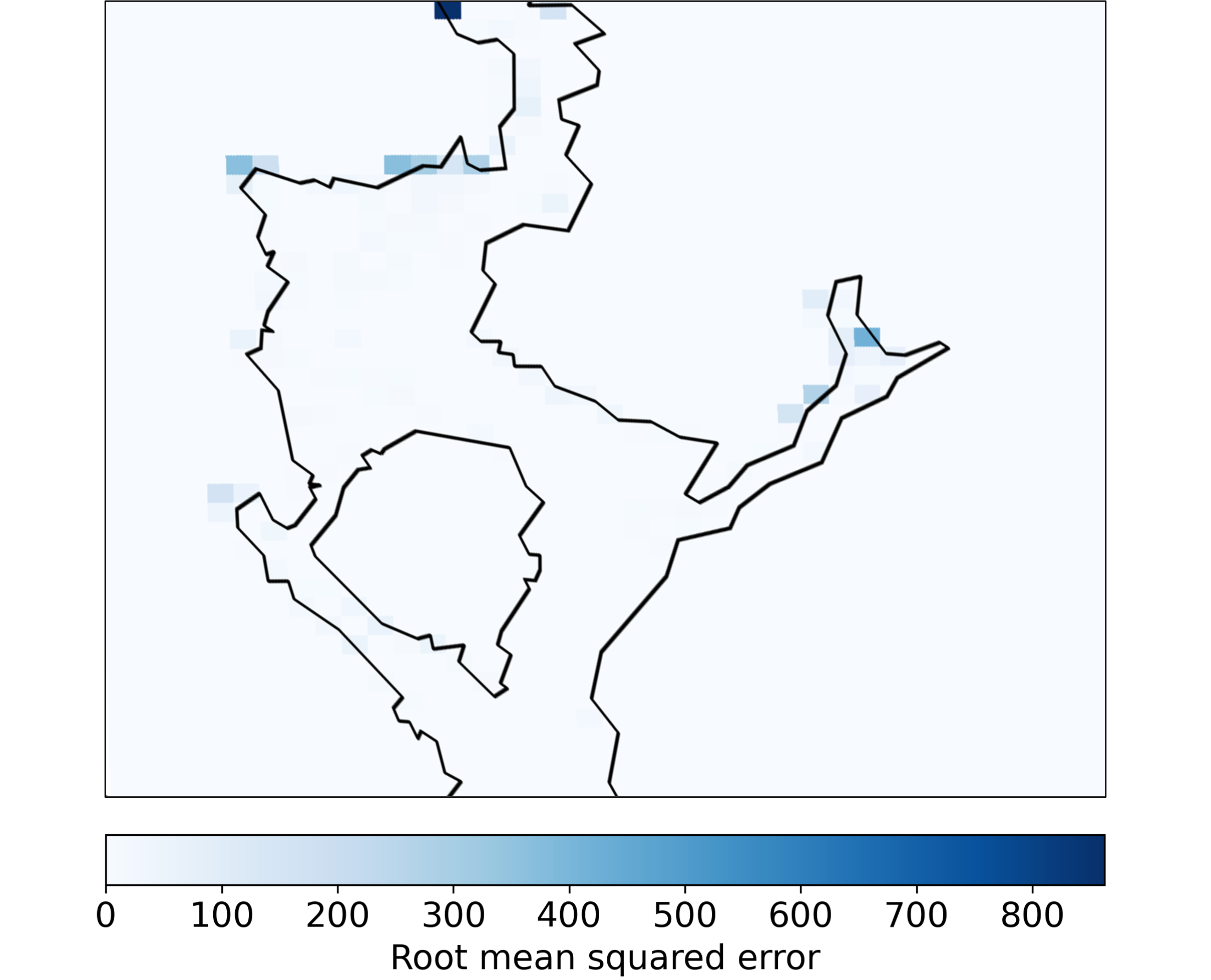}
         \caption{RMSE between (a) and (b)}
         \label{fig_RMSE}
    \end{subfigure}
    \caption{The sea lice infestation map forecast out to 7 days using (a) OSLM and (b) SALT. (c) shows the RMSE for each node between the two methodologies (a) and (b). The number of particles at a specific grid point is indicated by the color intensity, where darker color denotes higher numbers and lighter color denotes lower numbers of sea lice. As shown in (c), RMSE for areas of high infestation (nodes with +600 sea lice particles in (a) and (b)) show relatively low degree of error. This indicates that SALT can accurately predict areas of high infestation areas predicted by OSLM.}
    \label{fig:three graphs}
\end{figure*}

\begingroup
\setlength{\tabcolsep}{2pt} 
\renewcommand{\arraystretch}{1} 
\begin{table}[!t]
    \small
    \centering
    \resizebox{0.85\columnwidth}{!}{
        \begin{tabular}{l|c|ccccc}  
            \toprule
            \multirow{3}{*}{}
            & OSLM & \multicolumn{5}{c}{SALT} \\[-\rowheight] & & 
            \multicolumn{5}{c}{} \\ & & 
            \multicolumn{5}{c}{\% reduction using K-means} \\ &
             & 0.5\% & 5\%  & 10\% & 25\% & 40\%   \\  \hline
            \multicolumn{1}{r|}{\begin{tabular}[c]{@{}r@{}}Number of nodes\\[-3pt](thousands)\end{tabular}} & 289.0                    & 287.6   & 274.6 & 260.1 & 216.8 & 173.4  \\
            \multicolumn{1}{r|}{\begin{tabular}[c]{@{}r@{}}Ocean Model\\[-3pt]size (MB)\end{tabular}}       & 13.9                     & 12.7   & 11.9  & 11.1  & 8.7  & 6.3    \\
            \multicolumn{1}{r|}{\begin{tabular}[c]{@{}r@{}}K-means time\\[-3pt](mins)\end{tabular}}    & 0                     & 0.3   & 3.2  & 10.5 & 114.8 & 131.1  \\
            \toprule
        \end{tabular}%
    }
    \vspace{-1em}
    \caption{Performance metrics of the standard OSLM versus SALT for the Ocean Model.}
    \label{table_ocean_model}
\end{table}
\endgroup

\textbf{Ocean Model Comparison: } To assess the computational complexity of SALT in comparison to the conventional method, the size of the Ocean Model is used as the baseline. As shown in \autoref{table_ocean_model}, SALT achieves increasing reductions in computation complexity with decreasing number of initialized clusters, with a reduction of file size 1.8 times (12.7 MB model) to 11.6 times (6.3 MB model) smaller than the original Ocean Model data. This is achieved by using super-nodes derived from the unsupervised clustering result, where each super node represents one or more nodes in a local area. Initializing a low number of super nodes relative to the number of nodes in the original ocean model for clustering results in a significant data reduction, since the resulting number of nodes in the adaptive ocean model are lower. Similarly, initializing a high number of super nodes result in a smaller data reduction.

SALT's strong estimation accuracy is achieved by the reduction in spatiotemporal detail in low-complexity areas, such as land data points or water regions with a relatively homogeneous current speed, and the preservation of detail in high-complexity areas of the Ocean Model. \autoref{fig_num_cluster} shows an example of the number of clusters present in each spatial node. The number of clusters is computed by counting the number of unique clusters each node is associated with for the duration of the simulation. As seen, land nodes show a significantly lower number of clusters in comparison to water nodes (approx. 20 vs approx. 140 clusters).

\begingroup
\setlength{\tabcolsep}{2pt} 
\renewcommand{\arraystretch}{1} 
\begin{table}[!t]
\small
    \centering
    \resizebox{0.85\columnwidth}{!}{%
    \begin{tabular}{l|c|ccccc} 
    \toprule
    \multirow{3}{*}{} & OSLM & \multicolumn{5}{c}{SALT} \\
    [-\rowheight] & & \multicolumn{5}{c}{} \\
    & & \multicolumn{5}{c}{\% reduction using K-means} \\ &  & 0.5\% & 5\%  & 10\% & 25\% & 40\%   \\ 
    \hline \multicolumn{1}{r|}{\begin{tabular}[c]{@{}r@{}}Simulation time\\[-3pt](mins)\end{tabular}} & 3.4 & 3.4 & 3.4 & 3.4 & 3.4 & 3.4            \\
    \multicolumn{1}{r|}{RMSE}                                                            
    & 0 & 26.1  & 23.5 & 19.8 & 17.4  & 16.1   \\
    \toprule
    \end{tabular}%
    }
    \vspace{-.5em}
    \caption{Performance metrics of the standard OSLM versus SALT for the Dispersion Model.}
    \label{table_disp_model}
\end{table}
\endgroup

\begingroup
\setlength{\tabcolsep}{2pt} 
\renewcommand{\arraystretch}{1} 
\begin{table}[!t]
\small
    \centering
    \resizebox{0.85\columnwidth}{!}{%
    \begin{tabular}{l|c|ccccc} 
\toprule
\multirow{3}{*}{}
& OSLM & \multicolumn{5}{c}{SALT} \\[-\rowheight] & & 
\multicolumn{5}{c}{} \\ & & 
\multicolumn{5}{c}{\% reduction using K-means} \\ &
 & 0.5\% & 5\%  & 10\% & 25\% & 40\%   \\  \hline
\multicolumn{1}{r|}{\begin{tabular}[c]{@{}r@{}}Simulation time\\[-3pt](mins)\end{tabular}} & 3.4 & 3.4 & 3.4 & 3.4 & 3.4 & 3.4            \\
\multicolumn{1}{r|}{\begin{tabular}[c]{@{}r@{}}K-means time\\[-3pt](mins)\end{tabular}} & 0.0 & 0.3 & 3.2 & 10.5 & 114.8 & 131.1            \\
\multicolumn{1}{r|}{\begin{tabular}[c]{@{}r@{}}Total time\\[-3pt](mins)\end{tabular}} & 3.4 & 3.7 & 6.6 & 13.9 & 118.2 & 134.5            \\
\toprule
    \end{tabular}%
    }
    \vspace{-.5em}
    \caption{Time taken to run the individual components as well as total time taken to run the entirety of the standard OSLM versus SALT for the Dispersion Model.}
    \label{table_times}
\end{table}
\endgroup

\textbf{Dispersion Model Comparison: } The estimation accuracy of SALT is assessed through calculation of root-mean-square-error (RMSE) between the sea lice infestation pressure maps produced by OSLM and SALT. The maps show a count of the number of sea lice particles in each node, as shown in \autoref{fig_OSLM} and \autoref{fig_SALT}. RMSE is computed by aggregating the error between the nodes of the two methods. For sea lice infestation pressure models, it is important for the model to accurately model nodes of high sea lice densities. As shown in \autoref{table_disp_model}, SALT on average maintains strong estimation accuracy showing an RMSE value well below nodes with high sea lice density (800 particles/node), despite the reduction in the number of data points.

Furthermore, it can be seen that increasing the number of super-nodes decreases RMSE. Specifically, \autoref{fig_RMSE} shows an RMSE map of sea lice infestation pressure between the conventional method and SALT. Overall, RMSE is low across the entire map. In addition, SALT is able to accurately simulate most regions of high infestation pressure using significantly lower number of nodes. This is important as identifying areas of high infestation pressure is critical in sea lice control and management.

However, this decrease in RMSE comes at the cost of increasing the total compute time, as shown in \autoref{table_times}. This is due to the additional time taken by the clustering method, which increases with increase in clustering. Nevertheless, this does not pose a critical overhead since these processes are run offline once a week.

\section{Conclusions and Future Work}
We propose SALT, an efficient representation of the Ocean Model through the generation of adaptive lattice grids using unsupervised clustering. This technique can be used to overcome computational limitations to enable efficient dispersion estimation of local systems, which typically require much higher data storage requirements. To this end, our hope is that SALT enables a more comprehensive analysis of waterways to control sea lice in a rapidly changing climate and warming waters.

One known limitation of SALT is that by reducing the Ocean Model data size requirements, additional compute time is added to derive the adaptive Ocean Model. Future work would involve assessing and/or improving alternative unsupervised learning algorithms for increased clustering efficiency. Furthermore, the adaptive lattice can be leveraged to explore variable time stepping in the particle tracking model to improve computational efficiency while maintaining estimation accuracy.

\bibliography{SALT}

\begin{thebibliography}{16}
\providecommand{\natexlab}[1]{#1}
\providecommand{\url}[1]{\texttt{#1}}
\expandafter\ifx\csname urlstyle\endcsname\relax
  \providecommand{\doi}[1]{doi: #1}\else
  \providecommand{\doi}{doi: \begingroup \urlstyle{rm}\Url}\fi

\bibitem[Abraham et~al.(2013)Abraham, Baringer, Bindoff, Boyer, Cheng, Church,
  Conroy, Domingues, Fasullo, Gilson, Goni, Good, Gorman, Gouretski, Ishii,
  Johnson, Kizu, Lyman, Macdonald, Minkowycz, Moffitt, Palmer, Piola,
  Reseghetti, Schuckmann, Trenberth, Velicogna, and
  Willis]{abraham_review_2013}
Abraham, J.~P., Baringer, M., Bindoff, N.~L., Boyer, T., Cheng, L.~J., Church,
  J.~A., Conroy, J.~L., Domingues, C.~M., Fasullo, J.~T., Gilson, J., Goni, G.,
  Good, S.~A., Gorman, J.~M., Gouretski, V., Ishii, M., Johnson, G.~C., Kizu,
  S., Lyman, J.~M., Macdonald, A.~M., Minkowycz, W.~J., Moffitt, S.~E., Palmer,
  M.~D., Piola, A.~R., Reseghetti, F., Schuckmann, K., Trenberth, K.~E.,
  Velicogna, I., and Willis, J.~K.
\newblock A review of global ocean temperature observations: Implications for
  ocean heat content estimates and climate change.
\newblock \emph{Reviews of Geophysics}, 51\penalty0 (3):\penalty0 450--483,
  2013.

\bibitem[Amundrud \& Murray(2009)Amundrud and Murray]{amundrud_modelling_2009}
Amundrud, T.~L. and Murray, A.~G.
\newblock Modelling sea lice dispersion under varying environmental forcing in
  a scottish sea loch.
\newblock \emph{Journal of Fish Diseases}, 32\penalty0 (1):\penalty0 27--44,
  2009.

\bibitem[Asplin et~al.(2014)Asplin, Johnsen, Sandvik, Albretsen, Sundfjord,
  Aure, and Boxaspen]{asplin_dispersion_2014}
Asplin, L., Johnsen, I.~A., Sandvik, A.~D., Albretsen, J., Sundfjord, V., Aure,
  J., and Boxaspen, K.~K.
\newblock Dispersion of salmon lice in the hardangerfjord.
\newblock \emph{Marine Biology Research}, 10\penalty0 (3):\penalty0 216--225,
  2014.
\newblock ISSN 1745-1000.

\bibitem[Asplin et~al.(2020)Asplin, Albretsen, Johnsen, and
  Sandvik]{asplin_hydrodynamic_2020}
Asplin, L., Albretsen, J., Johnsen, I.~A., and Sandvik, A.~D.
\newblock The hydrodynamic foundation for salmon lice dispersion modeling along
  the norwegian coast.
\newblock \emph{Ocean Dynamics}, 70\penalty0 (8):\penalty0 1151--1167, 2020.

\bibitem[Barrett et~al.(2020)Barrett, Oppedal, Robinson, and
  Dempster]{barrett_prevention_2020}
Barrett, L.~T., Oppedal, F., Robinson, N., and Dempster, T.
\newblock Prevention not cure: a review of methods to avoid sea lice
  infestations in salmon aquaculture.
\newblock \emph{Reviews in Aquaculture}, 12\penalty0 (4):\penalty0 2527--2543,
  2020.

\bibitem[Bateman et~al.(2016)Bateman, Peacock, Connors, Polk, Berg, Krkosek,
  and Morton]{wild_salmon_damage_due_to_control_failure}
Bateman, A.~W., Peacock, S.~J., Connors, B., Polk, Z., Berg, D., Krkosek, M.,
  and Morton, A.
\newblock Recent failure to control sea louse outbreaks on salmon in the
  broughton archipelago, british columbia.
\newblock \emph{Canadian Journal of Fisheries and Aquatic Sciences},
  73\penalty0 (8):\penalty0 1164--1172, 2016.

\bibitem[Costello(2006)]{costello_ecology_2006}
Costello, M.~J.
\newblock Ecology of sea lice parasitic on farmed and wild fish.
\newblock \emph{Trends in Parasitology}, 22\penalty0 (10):\penalty0 475--483,
  2006.

\bibitem[Ford \& Myers(2008)Ford and Myers]{ford_global_2008}
Ford, J.~S. and Myers, R.~A.
\newblock A global assessment of salmon aquaculture impacts on wild salmonids.
\newblock \emph{PLoS Biology}, 6\penalty0 (2):\penalty0 e33, 2008.

\bibitem[Godwin et~al.(2020)Godwin, Fast, Kuparinen, Medcalf, and
  Hutchings]{warming_decreases_salmon_survival}
Godwin, S.~C., Fast, M.~D., Kuparinen, A., Medcalf, K.~E., and Hutchings, J.~A.
\newblock Increasing temperatures accentuate negative fitness consequences of a
  marine parasite.
\newblock \emph{Scientific Reports}, 10\penalty0 (1):\penalty0 18467, 2020.

\bibitem[MOWI(2019)]{mowi_salmon_2019}
MOWI.
\newblock Salmon farming industry handbook.
\newblock pp.\  114, 2019.

\bibitem[Myksvoll et~al.(2018)Myksvoll, Sandvik, Albretsen, Asplin, Johnsen,
  Karlsen, Kristensen, Melsom, Skardhamar, and
  {\AA}dlandsvik]{myksvoll_evaluation_2018}
Myksvoll, M.~S., Sandvik, A.~D., Albretsen, J., Asplin, L., Johnsen, I.~A.,
  Karlsen, O., Kristensen, N.~M., Melsom, A., Skardhamar, J., and
  {\AA}dlandsvik, B.
\newblock Evaluation of a national operational salmon lice monitoring
  system—from physics to fish.
\newblock \emph{PLOS ONE}, 13\penalty0 (7), 2018.

\bibitem[Pike \& Wadsworth(1999)Pike and Wadsworth]{pike_sealice_1999}
Pike, A.~W. and Wadsworth, S.~L.
\newblock Sealice on salmonids: Their biology and control.
\newblock In \emph{Advances in Parasitology}, volume~44, pp.\  233--337.
  Academic Press, 1999.

\bibitem[Press et~al.(2007)Press, H, Teukolsky, Vetterling, A, and
  Flannery]{press_numerical_2007}
Press, W., H, W., Teukolsky, S., Vetterling, W., A, S., and Flannery, B.
\newblock \emph{Numerical Recipes 3rd Edition: The Art of Scientific
  Computing}.
\newblock Cambridge University Press, 2007.
\newblock ISBN 978-0-521-88068-8.

\bibitem[Sandvik et~al.(2014)Sandvik, Bjørn, {\AA}dlandsvik, Asplin,
  Skar{\dh}hamar, Johnsen, Myksvoll, and Skogen]{sandvik_toward_2016}
Sandvik, A., Bjørn, P., {\AA}dlandsvik, B., Asplin, L., Skar{\dh}hamar, J.,
  Johnsen, I., Myksvoll, M., and Skogen, M.
\newblock Toward a model-based prediction system for salmon lice infestation
  pressure.
\newblock \emph{Aquaculture Environment Interactions}, 8:\penalty0 527--542,
  2014.
\newblock ISSN 1869-215X, 1869-7534.

\bibitem[Shephard \& Gargan(2021)Shephard and Gargan]{wild_salmon_damage}
Shephard, S. and Gargan, P.
\newblock Wild atlantic salmon exposed to sea lice from aquaculture show
  reduced marine survival and modified response to ocean climate.
\newblock \emph{{ICES} Journal of Marine Science}, 78\penalty0 (1):\penalty0
  368--376, 2021.

\bibitem[Torrissen et~al.(2013)Torrissen, Jones, Asche, Guttormsen, Skilbrei,
  Nilsen, Horsberg, and Jackson]{torrissen_salmon_2013}
Torrissen, O., Jones, S., Asche, F., Guttormsen, A., Skilbrei, O.~T., Nilsen,
  F., Horsberg, T.~E., and Jackson, D.
\newblock Salmon lice - impact on wild salmonids and salmon aquaculture.
\newblock \emph{Journal of Fish Diseases}, 36\penalty0 (3):\penalty0 171--194,
  2013.

\end{thebibliography}
\bibliographystyle{icml2021}

\end{document}